# OTFS - ORTHOGONAL TIME FREQUENCY SPACE

## A Novel Modulation Technique meeting 5G High Mobility and Massive MIMO Challenges

### Authors

Anton Monk, Ronny Hadani, Michail Tsatsanis, Shlomo Rakib

All with Cohere Technologies

### Abstract

In this paper we introduce a new 2D modulation technique called OTFS (Orthogonal Time Frequency & Space) that transforms information carried in the Delay-Doppler coordinate system to the familiar time-frequency domain utilized by traditional modulation schemes such as OFDM, CDMA and TDMA. OTFS converts the fading, time-varying wireless channel into a non-fading, time-independent interaction revealing the underlying geometry of the wireless channel. In this new formulation, all QAM symbols experience the same channel and all Delay-Doppler diversity branches of the channel are coherently combined. Reference signal multiplexing is done in the time-independent Delay-Doppler domain, achieving high density pilot packing, which is a crucial requirement for Massive MIMO. Regardless of the Doppler scenario, OTFS enables approaching channel capacity through linear scaling of throughput with MIMO order, thus realizing the full promise of Massive MIMO throughput gains even in challenging 5G deployment settings.

# 1.INTRODUCTION

The 5G air interface and associated waveform have to support a number of diverse requirements and usage scenarios, as described for example in [1-3] and in the NGMN[1] white paper on 5G [4]. One of the most challenging set of requirements in 5G is driven by the enhanced mobile broadband (eMBB) usage scenario and associated high speed deployment scenarios identified in 3GPP, for which current solutions are not well suited.

Of particular relevance to this paper are scenarios involving high speed traffic and vehicle-to-vehicle or vehicle-to-infrastructure (V2X). These scenarios include proposed vehicle speeds of up to 500 km/h. It is well known that under higher Doppler conditions, channel estimation performance and associated OFDM modulation performance break down. This performance degradation is exacerbated with higher order MIMO due to the strong correlation between performance and accurate channel estimation [5,6]. In addition, capacity-increasing techniques such as cooperative multipoint and massive MIMO require accurate channel estimation and the support for a large number of reference signals under all Doppler conditions in order to approach their promised performance gains.

All of the deployment scenarios associated with eMBB are characterized by large numbers of antenna elements (up to 256 currently proposed in 3GPP). Leveraging these antenna elements in a massive MIMO architecture requires a highly efficient reference signal multiplexing mechanism, otherwise the bandwidth overhead required for the large number of antenna ports will overwhelm the available data bandwidth.

Here we propose a new modulation scheme and a new reference signal architecture, which both provide significant performance improvements over existing methods in high Doppler environments and for high-order MIMO systems. These techniques also greatly enhance reference signal multiplexing efficiency relative to existing solutions.

The new modulation scheme we propose is OTFS (Orthogonal Time Frequency Space), which modulates each information (e.g., QAM) symbol onto one of a set of two dimensional (2D) orthogonal basis functions that span the bandwidth and time duration of the transmission burst or packet. The modulation basis function set is specifically derived to directly represent the dynamics of the time- varying multipath channel. OTFS can be implemented as a pre- and post-processing block to filtered OFDM systems, thus enabling architectural compatibility with LTE.

OTFS transforms the time-varying multipath channel into a time-independent two dimensional channel in the Delay-Doppler domain that directly represents the geometry of the various reflectors composing the wireless link. In this way, OTFS eliminates the difficulties in tracking time-varying fading, particularly in high speed vehicle communications. Due to its ability to extract the full diversity of channel across time and frequency, OTFS enables linear scaling of throughput with the number of antennas in moving vehicle applications. In addition, since the Delay-Doppler channel representation is very compact, OTFS enables dense and flexible packing of reference signals, a key requirement to support the large antenna arrays used in massive MIMO applications.

---

[1] Next Generation Mobile Networks is an alliance of global operators and vendors

The remainder of this paper is organized as follows. In Section 2 we describe the wireless multipath channel in terms of the Delay-Doppler representation used by OTFS. Section 3 provides the details of OTFS modulation, including its underlying premise, basis functions, signal processing steps, and receiver design. Performance results demonstrating the benefits of OTFS over existing methods are provided in Section 4. Finally, in Section 5, we summarize the OTFS technique and its benefits for high-mobility and massive MIMO 5G networks.

# 2. THE WIRELESS CHANNEL

## 2.1.   THE TIME VARYING IMPULSE RESPONSE

The multipath fading channel is commonly modeled based on its time varying impulse response, $\hbar(\tau, t)$. In wireless OFDM systems like LTE, the channel is equivalently modeled as a time-varying frequency response, $H(f, t)$. In LTE this frequency response is typically estimated every OFDM symbol time in order to equalize the channel.

Either of these Time-Frequency domain representations, while general, does not give us much insight into the behavior and variations of the channel other than the fact that higher mobility results in faster variations of the multipath components. Since constructive and destructive addition of these multipath components causes signal fading, faster variation of these components leads to more rapid fluctuations in the channel. The frequency response rate of variation is also proportional to the signal carrier frequency. Thus, the faster the reflector, transmitter, and/or receiver velocity and the higher the frequency band, the more rapidly changes in the channel frequency response occur.

## 2.2.   THE DELAY-DOPPLER IMPULSE RESPONSE

A more useful and insightful channel model, which is also commonly used in radar theory, is the Delay-Doppler impulse response. In this representation, the received signal is a superposition of reflected copies of the transmitted signal, where each copy is time-delayed by the path delay $\tau$ (as in $\hbar(\tau, t)$ above), and frequency shifted by the Doppler shift $\nu$ (related to the speed of the reflected copy) and weighted by the time-independent Delay-Doppler impulse response $h(\tau, \nu)$ for that $\tau$ and $\nu$. In this representation, $h(\tau, \nu)$ directly represents the geometry of the mobile channel. In addition to the intuitive nature of this representation, the Delay-Doppler impulse response maintains the generality of the time-varying impulse response. In other words, it can represent complex Doppler trajectories, like accelerating vehicles, reflectors, etc. In fact, the Delay-Doppler impulse response is related to the time-varying frequency response through a relationship known in the math community as a 2D Symplectic Fourier Transform [7].

Any of the above channel representations suffice to fully describe the wireless multipath channel. However, the Delay-Doppler representation has some distinct features that are interesting and applicable in high mobility environments. In particular, the Delay-Doppler impulse response changes far more slowly than its time-varying counterpart since the channel geometry changes very slowly relative to the communication timescale. For example, consider a 3.6 GHz link to a car traveling at 100 km/h – this translates to approximately 300 Hz of Doppler. As the channel coherence time in the Time-Frequency domain is the inverse of its Doppler, the impulse response for this channel varies rapidly over a fraction of a millisecond – so rapidly, in fact, that for such channels in an LTE/OFDM system there is not sufficient time to estimate and feed back the channel; Hence, open-loop techniques must be used. In contrast, the Delay-Doppler domain view of the channel is time-invariant over a large observation time (both the distance and velocity of a moving reflector are essentially unchanged over around 10 msec, for example), because the Delay-Doppler impulse response mirrors the physical geometry of the channel.

### 2.3. OTFS BASIS FUNCTIONS

A modulation scheme is defined by the underlying basis functions that comprise the waveform. Typically, each modulation symbol (generally QAM information symbols) is assigned a different orthogonal basis function. In OFDM, these basis functions are one-dimensional orthogonal frequency-domain sinusoids. In TDMA, or single carrier modulation, the basis functions are one dimensional orthogonal time-domain filtered pulses. And in CDMA, the basis functions are orthogonal one-dimensional time-domain codes. In all three of these well-known modulation schemes, QAM symbols are assigned to basis functions in the Time-Frequency domain.

OTFS, in contrast, is a completely novel modulation scheme comprised of two-dimensional basis functions that are orthogonal to both translation and modulation, i.e., to time and frequency shifts – the two primary features of the time-varying wireless channel. As will be explained in more detail in the next section, in OTFS QAM symbols are indexed by points on a lattice or grid in the Delay-Doppler domain, not the Time-Frequency domain and, through the two dimensional OTFS Transform, each QAM symbol weights a 2D basis function in the Time-Frequency domain. The choice of these unique orthogonal basis functions renders the effects of the wireless channel truly time-independent in the Delay-Doppler domain.

# 3. OTFS MODULATION

OTFS modulation has numerous benefits that tie into the challenges that 5G systems are trying to overcome. Arguably, the biggest benefit and the main reason to study this modulation is its ability to transform a channel that randomly fades within the time-frequency frame into a stationary, deterministic and non-fading channel between the transmitter and the receiver. As will be seen, in the OTFS Delay-Doppler domain all information symbols experience the same static channel response and the same SNR.

Further, OTFS best utilizes the fades and power fluctuations in the received signal to maximize capacity. To illustrate this point, assume that the channel consists of two reflectors which introduce peaks and valleys in the channel response across time, across frequency, or across both. An OFDM system can theoretically address this problem by allocating power resources according to the waterfilling principle [8]. However, due to practical difficulties related to the speed of channel fluctuations, such approaches are challenging to implement in high-mobility OFDM systems. In particular, poor channel estimates in such environments leads to mismatched water-filling, such that parts of the time-frequency frame having excess received energy, while other parts have too low received energy. An OTFS system resolves the two reflectors and the receiver equalizer coherently combines the energy of the two reflectors, providing a non-fading channel with the same SNR for each symbol. It therefore provides a channel that is designed to maximize capacity under the transmit assumption of equal power allocation across symbols (which is common in existing wireless systems), using only standard AWGN codes. In other words, OTFS does not require closed-loop adaptation to the channel as the channel in the Delay-Doppler domain is relatively static, hence the overall system is more robust and with lower complexity than existing adaptive OFDM implementations.

In addition, OTFS operates in a domain in which the channel can be characterized in a very compact form. This has significant implications for addressing the channel estimation bottlenecks that plague current multi-antenna systems and can be a key enabling technology for addressing similar problems in future massive MIMO systems.

One key benefit of OTFS is its ability to easily handle extreme Doppler channels. This is not only useful in vehicle-to-vehicle, high speed train and other 5G applications that are Doppler intensive, but can also be an enabling technology for mm-Wave systems where Doppler effects will be significantly amplified.

Last, but not least, the compact channel estimation process that OTFS provides can be essential to the successful deployment of advanced technologies like Cooperative Multipoint (Co-MP) and distributed interference mitigation or network MIMO, that require accurate channel estimation for large numbers of antenna ports in any mobility scenario [9].

## 3.1. OTFS MODULATION PRINCIPLE

OTFS works in the Delay-Doppler coordinate system using a set of basis functions orthogonal to both time and frequency shifts. Both data and reference signals or pilots are carried in this coordinate system. The Delay-Doppler domain mirrors the geometry of the wireless channel, which changes far more slowly than the phase changes experienced in the rapidly varying time-frequency domain. OTFS symbols experience the full diversity of the channel over time and frequency, trading latency for performance in high Doppler scenarios.

Figure 1 illustrates the modulation and demodulation steps as well as the channel effects on the modulation. The transmit information symbols (QAM symbols) are placed on a lattice or grid in the 2 dimensional Delay-Doppler domain (shown top left in green) and transformed to the Time-Frequency domain through a two dimensional Symplectic Fourier Transform. Recall that this is the familiar Time-Frequency domain where OFDM QAM symbols reside. In contrast, in OTFS, each QAM symbol is spread throughout this Time-Frequency plane (i.e., across the selected signal bandwidth and symbol time) utilizing a different basis function. As a result, all symbols of the same power have the same SNR and experience exactly the same channel. The implication is that, given the appropriate frequency and time observation window, there is no frequency or time selective fading of QAM symbols. Subsequently, the signal is passed through a multicarrier filter bank, allowing for the same filter shaping benefits seen in various forms of filtered OFDM. At the receive side, the inverse processing is performed.

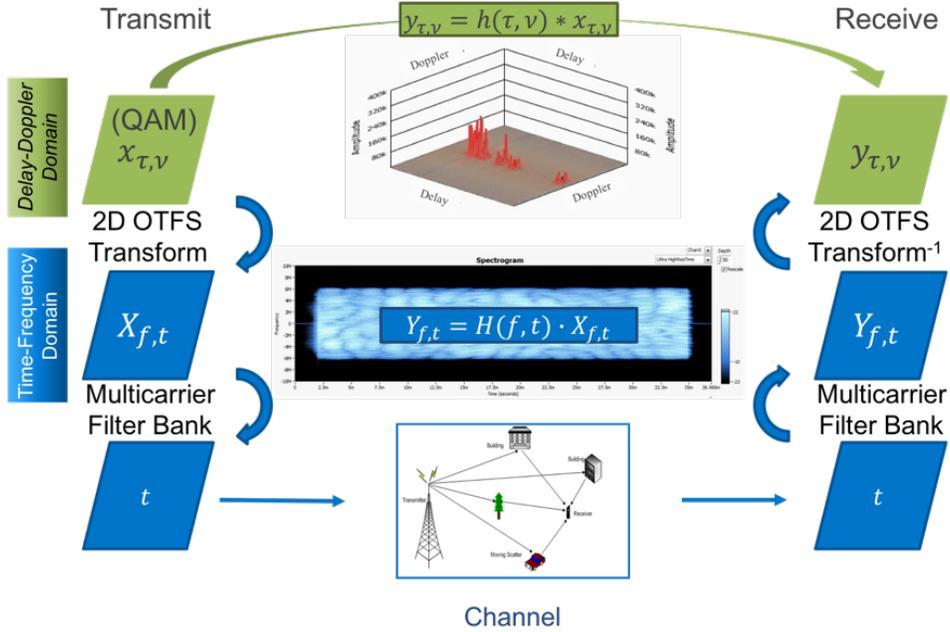

*Figure 1: OTFS Processing*

Figure 1 also shows the interaction of the channel with the transmit signal. The lower graphic shows the physical nature of the wireless channel which consists of transmitter, receiver and various reflectors. The center graphic shows the time-varying frequency response of a high speed, high delay spread (300 Hz Doppler, 5 $\mu$sec) channel displayed as a spectrogram[2]. This complex channel can be seen to be both highly frequency and highly time selective. In this domain, the relationship between the transmit and receive signals is multiplicative.

The top graphic shows the Delay-Doppler domain of the same channel. In this domain, the channel is observed over a longer observation time and represented by a compact Delay-Doppler impulse response. The OTFS Symplectic transform converts the multiplicative action of the channel into a 2D convolutive interaction with the transmitted QAM symbols.

To summarize, in OTFS information symbols are indexed by points on a lattice or grid in the Delay-Doppler domain. Through the OTFS Symplectic Transform each QAM symbol weights a 2D basis function defined in the Time-Frequency domain. The frequency domain samples at each time are transformed into time domain waveforms using filter banks.

## 3.2.   OTFS PROCESSING STEPS

In OTFS, the information QAM symbols are arranged over a $N_t x N_f$ grid on the Delay-Doppler plane. A 2D Symplectic Fourier Transform translates every point on this Delay-Doppler plane into a corresponding basis function that covers the entire Time-Frequency plane (the basis functions are 2D orthogonal complex exponentials). The Symplectic Fourier Transform differs from the more well-known Cartesian Fourier Transform in that the exponential functions across each of the two dimensions have opposing signs and the coordinates are flipped in the two domains. This is necessary as it matches the behavior of the Delay-Doppler channel representation relative to the time-varying frequency response representation of the channel, as explained above.

---

[2] A spectrogram is the time-varying frequency response of the channel with brighter areas representing lower attenuation and darker areas representing greater attenuation.

Thus OTFS QAM symbols are transformed onto an identically-sized grid representing sample points in the Time-Frequency domain and the energy of each QAM symbol is spread over the Time-Frequency domain. Recall that this is the same grid over which OFDM QAM symbols (or any of its filtered multi-carrier variants) are defined ($N_f$ is equivalent to the number of subcarriers and $N_t$ to the number of multi-carrier symbols). The mapping of information symbols to the time-frequency basis functions is illustrated in Figure 2.

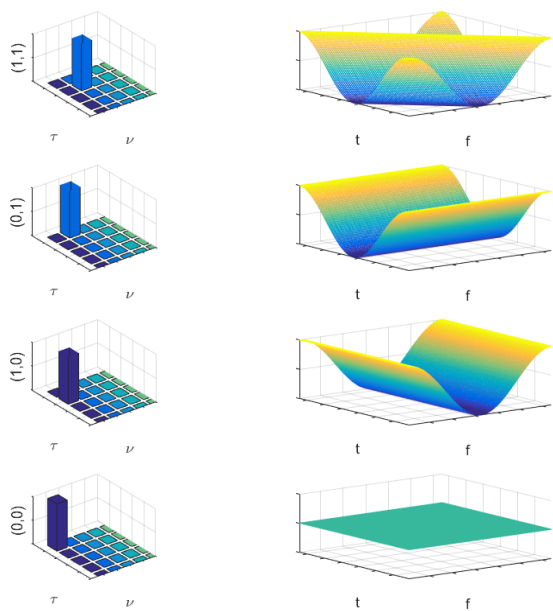

*Figure 2: Mapping of Information Symbols (delay-Doppler Domain) to Basis Functions (Time-Frequency Domain)*

By its very nature, OTFS enables every packet, regardless of size, to capture the full diversity of the channel since all QAM symbols are spread over the observed frequency and time block. This has valuable implications for other 5G requirements such as deep in-building penetration to IoT devices and sensor networks, as described in [1]. These use cases require 20 dB or more additional processing gain. OTFS enables small packets and data rates to be supported with higher power, i.e., a seamless tradeoff of data rate, power and processing gain, while still reaping the same benefits of capturing all delay and Doppler diversity branches in the channel.

### 3.3.    OTFS RECEIVER

In terms of implementation, the OTFS transform consists of pre- and post-processing blocks to the OFDM modulator and demodulator in the transmitter and receiver respectively, as depicted in Figure 1. This is analogous to the pre- and post-processing DFT blocks used to implement SC-FDMA on top of an underlying OFDM signal chain in LTE. It should be noted that the OTFS modulation can also be derived as a pre- and post-processing of multicarrier systems other than OFDM (e.g., filter bank multicarrier);

### 3.4. OTFS REFERENCE SIGNALS

5G solutions are expected to include massive MIMO antenna arrays of up to 256 antenna elements. These require a large number of reference signals to be multiplexed in order to simultaneously estimate the channels to and from the UEs. The reference signal overhead is a crucial component in determining the overall throughput of the link.

Reference signals (RSs) are typically placed in the Time-Frequency domain to assist the receiver in estimating the channel, generally in a coarse (regular or irregular) grid, as in LTE. Multiple antenna ports are multiplexed on the same coarse grid using different (ideally orthogonal) signature sequences. However, this orthogonality is often lost after transmission through the channel. Understanding the effects of the channel in the Delay-Doppler domain can lead to significant improvements in RS multiplexing overhead when multiplexing antenna port RSs in the Delay-Doppler domain.

OTFS reference signals or pilots are carried in the Delay-Doppler domain as impulses to probe the channel. Each pilot has a space reserved around it to account for the maximum delay and Doppler spread of the channel. Like the information symbols, the pilots experience the same time and frequency diversity of the channel over the full observation bandwidth and time. The interaction with the channel results in a 2D convolution of the Delay-Doppler impulse response with the pilot – the effects of which are local, that is a delta in the Delay-Doppler domain will be spread only to the extent of the support of the channel in the delay and in the Doppler dimensions. This fact provides the blueprint for multiplexing antenna ports in this domain, i.e., represent each antenna port sequence as an RS impulse, and space the impulses sufficiently far apart so that when the impulses are spread by the channel they still do not overlap, or overlap minimally. Figure 3 shows an example of such an arrangement of RS antenna ports in the delay-Doppler domain. Notice that each antenna port RS in Figure 3 is generally affected by a different channel.

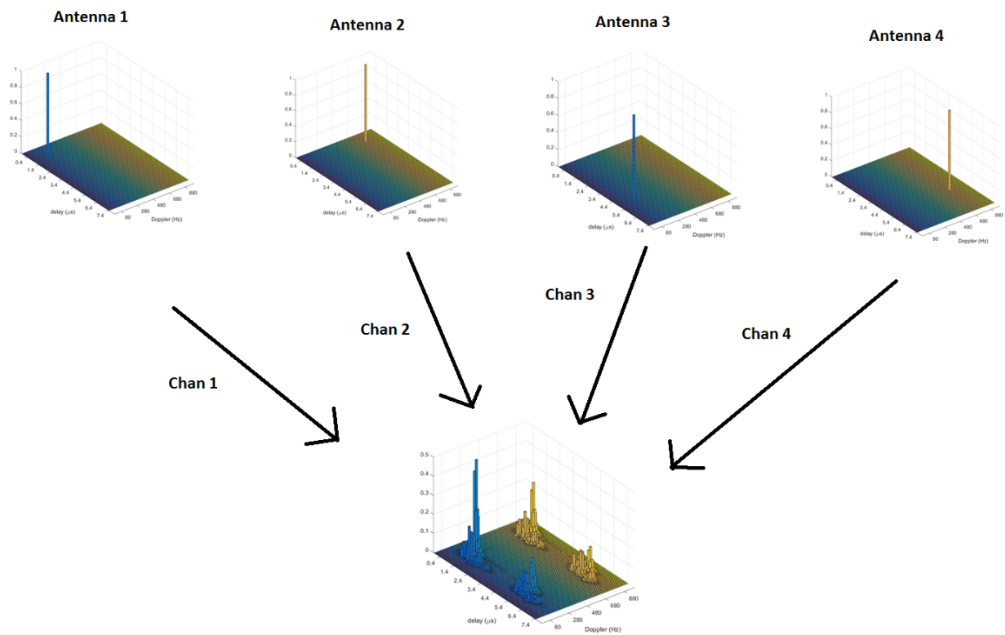

*Figure 3: Antenna Ports Multiplexed in the Delay-Doppler Domain*

The multiplexed reference signals are sampled in the Time-Frequency domain according to a selected coarse grid that does not overlap with the data grid points. This enables the observation window for estimating the channel from the reference signals to be independent of the data. Importantly, it also allows OTFS reference signals to be utilized for both OTFS as well as any multicarrier modulation, including OFDM and other proposed 5G waveforms.

Because the reference signals are multiplexed in the Delay-Doppler plane, which mirrors the geometry of the wireless channel, they can be very densely packed, based on the delay and Doppler characteristics of the channels. For example, it can be shown that for a 5 $\mu s$ delay spread and $100 Hz$ Doppler spread channel, 88 reference signals can be multiplexed in 7% of the available bandwidth for an overhead per antenna port of just 0.08%. In addition, further efficiency can be obtained with knowledge of channel conditions for different users or groups of users by flexibly assigning the users or groups different pilot spacing in the Delay-Doppler domain.

# 4. PERFORMANCE RESULTS

In this section we present some simulation results to illustrate the performance of an OTFS system. We compare with an OFDM LTE system with the same PHY parameters (10 MHz BW, 15 kHz subcarrier spacing, 1msec TTI) and the same Turbo FEC coding and assume a TM3 precoder for OFDM and no precoder for OTFS. We evaluate the performance in the 3GPP ETU-300 channel model (5 $\mu s$ delay spread, 300 Hz maximum Doppler). Results are shown for 1x1, 2x2, 4x4 and 8x8 MIMO.

The performance for both systems is compared in ideal conditions to provide an indication of the potential performance. Ideal channel estimation and genie-aided equalization are used for both OFDM and OTFS simulations. MMSE equalization with genie aided successive interference cancellation is used for OFDM based on the ideal channel, while MMSE-DFE is used for OTFS in the Delay-Doppler domain with no error propagation based on the ideal channel. We use the rate 1/3 Turbo code used in LTE for both systems but we relax the packet size constraints/quantization to get identical code rates for both systems. We also map one codeword per layer for both systems. We use the identity precoder for the OTFS simulation, the TM3 precoder for the OFDM 1x1, 2x2, and 4x4 simulations, and the identity precoder for the 8x8 simulation. The TM3 precoder randomizes the channel attenuation that each stream experiences across frequency bins resulting in improved performance for LTE systems. It was included in the simulation for all but the 8x8 case for which it has not been defined yet. The question of the optimal precoder for OTFS is an interesting one, but is outside the scope of this paper, hence only the identity precoder was used for OTFS, for all test cases.

The performance comparison is illustrated in Figure 4 in terms of throughput per stream versus SNR. For each SNR point, the throughput for all MCSs is estimated by simulation and the maximum of these throughputs is plotted. This plot is meant to capture the performance of an open loop MIMO system with slow adaptation where only the average channel quality index (CQI) is fed back to the transmitter. Notice the superior performance of OTFS, which grows as the MIMO order grows. In fact, OTFS performance is seen to scale linearly with MIMO order, even under high Doppler scenarios.

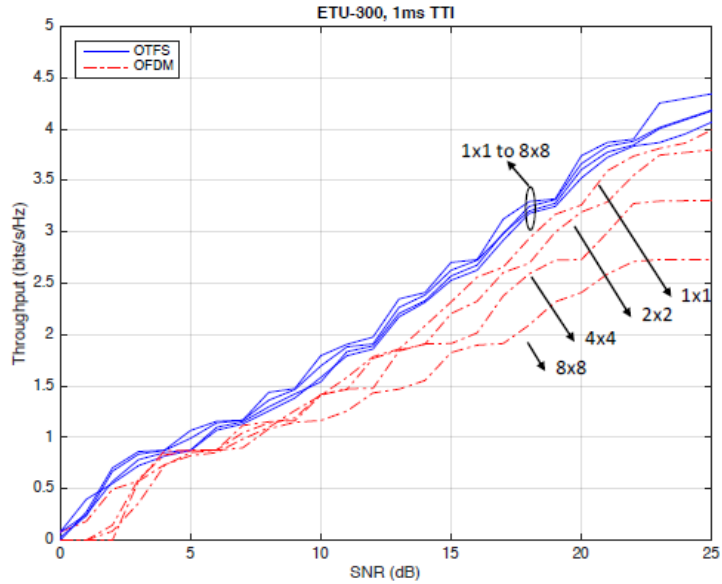

*Figure 4: Throughput per Stream OTFS vs OFDM for ETU-300 with 1x1, 2x2, 4x4 and 8x8 MIMO*

In the next simulation we study the performance of OTFS as a function of FEC codeblock size. Figure 5 shows BLER SISO performance for the EVA-600 channel, which represents a high-speed vehicle (250 km/h at 4 GHz), when the codeblock size is limited to 500, 1000 and 2000 bits. LTE CQI index 9, corresponding to modulation and coding of 16QAM and rate 0.6 FEC was used in this simulations (16 QAM, rate 0.6). Notice that the OTFS performance does not depend on the codeblock size, but the OFDM performance is negatively affected when the codeblock size gets smaller. This is because in the OTFS case all codeblocks experience the full diversity of the channel across the whole TTI, while in the OFDM case smaller codeblocks only occupy a small portion of the time-frequency plane and hence experience a smaller degree of diversity. In fact, in OTFS transmission every QAM symbol in the Delay-Doppler plane experiences the same SNR which is equal to the geometric mean of the channel SNR across time and frequency. This phenomenon is particularly pronounced in high Doppler scenarios, for example in the high speed train deployment scenario.

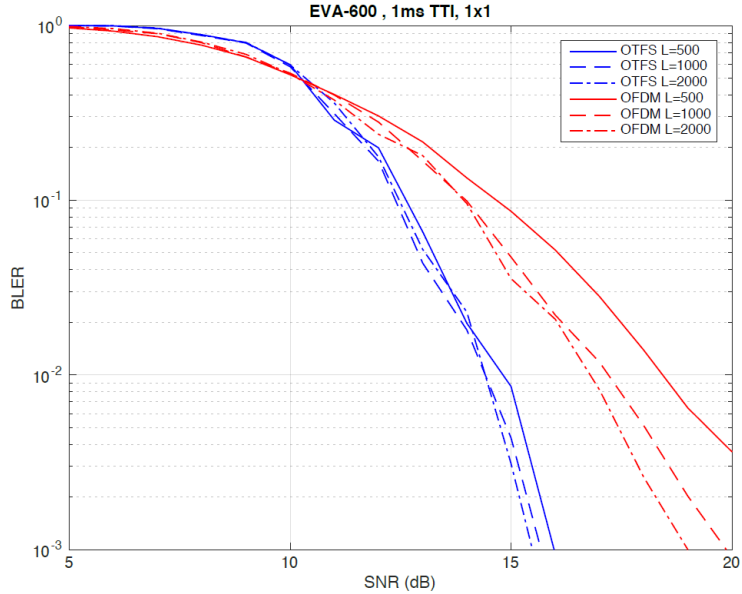

*Figure 5: BLER SISO Performance for Codeblock Size of 500, 1000, and 2000 Bits (EVA-600)*

Next, we evaluate the impact on performance of a non-genie-aided equalizer versus the idealized genie-aided equalizer assumed in the results above. We have chosen the genie-aided DFE (no error propagation) for the OTFS system in order to avoid a full analysis of various receiver architectures in terms of performance and complexity which is outside the scope of this paper (we have similarly used a genie-aided SIC for OFDM). However, the performance of the genie-aided DFE should not be assumed to be better than the performance of more advanced receivers without the genie-aided assumption. To illustrate this point, in Figure 6 we compare the performance of the genie-aided DFE with a non-genie-aided frequency domain turbo equalizer for a 2x2 and a 4x4 system. As can be seen in the figure, the non-genie-aided turbo equalizer outperforms the genie-aided DFE.

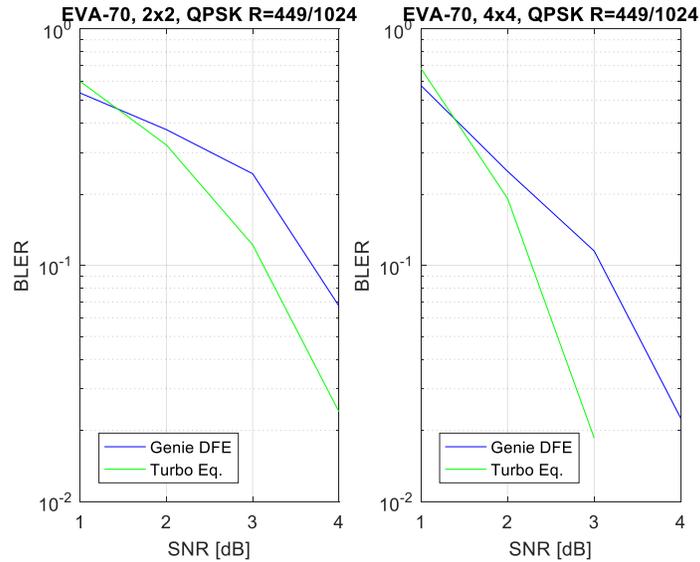

*Figure 6: BLER Performance of DFE vs Turbo Equalizer*

# 5. SUMMARY

OTFS is a new air interface paradigm with important spectral efficiency advantages in high order MIMO and high Doppler scenarios. OTFS also provides efficiency in pilot packing of reference signals for channel estimation and prediction. All reference signals and QAM symbols are carried in the Delay-Doppler domain and experience the same channel response over the transmission /observation interval and extract the maximum diversity of the channel in both time and frequency dimensions. This allows the FEC layer to operate on a signal with a uniform Gaussian noise pattern, regardless of the particular channel structure. OTFS has a natural architectural compatibility with OFDM, based on its underlying multicarrier components. Moreover, the reference signal architecture supports any form of multicarrier modulation.

3GPP has identified a variety of eMBB deployment scenarios that focus on high vehicle speed and massive MIMO antenna arrays. The new radio air interface must support high spectral efficiency in high Doppler environments while supporting a large number of antennas. OTFS is ideally suited for these requirements, providing: high spectral efficiency; accurate channel estimation and prediction; and very efficient and flexible reference signals for massive MIMO applications.

# 6. ACKNOWLEDGMENT

The authors would like to thank Prof. Andrea Goldsmith for reading the manuscript and providing valuable comments and suggestions.